\documentclass[conference]{IEEEtran}
\IEEEoverridecommandlockouts
\usepackage{cite}
\usepackage{amsmath,amssymb,amsfonts}
\usepackage{algorithmic}
\usepackage{graphicx}
\usepackage{textcomp}
\usepackage{bmpsize}
\usepackage{xcolor}
\usepackage[utf8]{inputenc}
\usepackage{textcomp}
\usepackage{newunicodechar}
\newunicodechar{−}{-}
\usepackage{balance}
\usepackage{tabularx}
\usepackage{booktabs}
\usepackage{comment}
\usepackage{dblfloatfix}
\usepackage{eso-pic}
\usepackage[colorlinks=false,urlcolor=black]{hyperref}
\def\BibTeX{{\rm B\kern-.05em{\sc i\kern-.025em b}\kern-.08em
    T\kern-.1667em\lower.7ex\hbox{E}\kern-.125emX}}
\begin{document}

\title{AutoGraphAD: Unsupervised network anomaly detection using Variational Graph Autoencoders}

\author{\IEEEauthorblockN{Georgios Anyfantis}
\IEEEauthorblockA{\textit{Department of Computer Architecture} \\
\textit{Universitat Politècnica de Catalunya}\\
Barcelona, Spain \\
georgios.anyfantis@upc.edu}
\and
\IEEEauthorblockN{Pere Barlet-Ros}
\IEEEauthorblockA{\textit{Department of Computer Architecture} \\
\textit{Universitat Politècnica de Catalunya}\\
Barcelona, Spain \\
pere.barlet@upc.edu}
}

\maketitle

\AddToShipoutPictureFG*{%
  \AtPageLowerLeft{%
    \put(54,28){%
      \parbox{0.85\paperwidth}{%
        \scriptsize
        \textcopyright~2026 IEEE. Personal use of this material is permitted.
        Permission from IEEE must be obtained for all other uses.
      }%
    }%
  }%
}

\begin{abstract}

Network Intrusion Detection Systems (NIDS) are essential tools for detecting network attacks and intrusions. While extensive research has explored the use of supervised Machine Learning for attack detection and characterisation, these methods require accurately labelled datasets, which are very costly to obtain. Moreover, existing public datasets have limited and/or outdated attacks, and many of them suffer from mislabelled data. To reduce the reliance on labelled data, we propose AutoGraphAD, a novel unsupervised anomaly detection approach based on a Heterogeneous Variational Graph Autoencoder. AutoGraphAD operates on heterogeneous graphs, made from connection and IP nodes that represent network activity. The model is trained using unsupervised and contrastive learning, without relying on any labelled data. The model's losses are then weighted and combined in an anomaly score used for anomaly detection. Overall, AutoGraphAD yields the same, and in some cases better, results than Anomal-E, but without requiring costly downstream anomaly detectors. As a result, AutoGraphAD achieves around 1.18 orders of magnitude faster training and 1.03 orders of magnitude faster inference, which represents a significant advantage for operational deployment.

\end{abstract}


\section{Introduction}
\label{sec:intro}

In recent years, attacks and intrusions have been a growing problem. The attacks reported each year have been growing exponentially, with attacks becoming more and more sophisticated \cite{9623451}. Thus, more robust Network Intrusion Detection Systems (NIDS) are needed to deal with the increasing volume and complexity of network attacks.

A substantial body of research has explored the use of Machine Learning (ML) to detect and characterise attacks \cite{8171733, LightweightIDS, 10127959, 10.1007/978-981-13-1921-1_35}. However, most of these approaches are susceptible to adversarial attacks and therefore do not suit real-life deployment \cite{7467366}. Graph Neural Networks (GNNs) are known to be more robust against adversarial attacks, as they use both network data and learn the structures and relationships among network traffic flows \cite{pujolperich2021unveilingpotentialgraphneural, 10123384, 10.1145/3664476.3664515}.

In the NIDS domain, the number of datasets that can be used to create and train ML models remains limited \cite{goldschmidt2025networkintrusiondatasetssurvey}. Many public NIDS datasets contain synthetic elements, outdated attacks, or labelling limitations, creating a distorted image of how attacks are structured and work \cite{PINTO2025111177}.

Labelling datasets is a costly and laborious task, which explains the scarcity of high-quality labelled data sets \cite{Braun_2024}. The main challenge is that each network flow must be examined and annotated by a domain expert, such as a Security Analyst \cite{GUERRA2022102810}. This makes labelling large real-world network data almost impossible, which is why most research in this area relies on synthetic datasets \cite{GUERRA2022102810}.

Anomaly detection is a good approach to mitigate reliance on labelled datasets, as we only need normal data to establish a baseline, and anything that deviates from it is considered anomalous. Anomal-E is arguably the most representative and influential approach in graph-based network anomaly detection \cite{Caville_2022}. 

In the case of Anomal-E and other approaches \cite{zoubir2024integratinggraphneuralnetworks, 10.1007/978-3-030-50423-6_12}, they rely on traditional downstream anomaly detection algorithms, which are often not suitable for a streaming environment \cite{app13106353}, making them very difficult to implement in real-world networks. This is an additional layer of computation that needs to be retuned often to account for concept drift \cite{Webb2016-vo} and the network's dynamic nature \cite{app13106353}. As both the model and the estimator would need to be taken offline to retrain and then redeploy.

Moreover, certain one class classification algorithms tend to become prohibitively expensive when depending on the size of the embedding used. This can be seen in the case of OneClassSVM \cite{NIPS1999_8725fb77} and its linear variants, where the dimensionality of the features negatively affects training and inference time, memory complexity, and overall performance \cite{scikit-learn}. Anomal-E \cite{Caville_2022} uses 256 vector embeddings, which are large and can negatively affect detection algorithms.

An interesting approach to avoid reliance on a downstream estimator is the use of dynamic Graph Autoencoders (GAEs) to detect anomalies. This approach has been used successfully in network monitoring as seen in GAT-AD \cite{LATIFMARTINEZ2025110830}, where the deviation of the predicted values from the actual values is used to detect anomalies.

In this paper, we propose AutoGraphAD, a novel approach to unsupervised anomaly detection based on Variational Graph Autoencoders (VGAEs). In our approach, we represent network traffic as a heterogeneous graph with two distinct types of nodes, IP and Connection nodes. We then performed unsupervised contrastive training for the VGAE, without requiring any labelled data. We rely solely on reconstruction errors and metrics to calculate the anomaly score, and we do not rely on any other downstream detectors. We have tested multiple different architectural variations to determine which implementation yields the best results at different levels of contamination (i.e., presence of anomalous flows) of the training dataset. The choice of testing at different contamination levels arises as we want to evaluate how robust the proposed architecture is. AutoGraphAD yields the same, and in some cases better, results than Anomal-E while not relying on the use of costly downstream estimators. Thus, faster training by 1.18 orders of magnitude and faster inference by 1.03 orders of magnitude is achieved, showcasing a significant advantage for operational deployment. We evaluate the method under multiple contamination levels and show comparable detection quality to Anomal-E while substantially reducing training and inference time.

\section{Related Work}

NIDS are a broad area with thousands of proposals; a review of all proposals is outside the scope of this paper. Thus, we will focus on related work in terms of GNN-based NIDS. Surveys on NIDS and GNN-based NIDS can be found here \cite{goldschmidt2025networkintrusiondatasetssurvey, ZHONG2024103821}.

GNN-based NIDS are mostly supervised. E-GraphSAGE \cite{Lo_2022} is a widely used representative baseline as it is a widely evaluated flow-graph GNN \cite{Venturi2023}. The model works by implementing E-GraphSAGE, a variation of GraphSAGE that works by processing only the Edge features rather than the node features that traditional GraphSAGE captures \cite{NIPS2017_5dd9db5e}. It is trained to perform edge (flow) classification. GraphIDS is a recently published model based on E-GraphSAGE \cite{guerra2025selfsupervisedlearninggraphrepresentations} by augmenting flow representation learning with a masked encoder-decoder Transformer. ARGANIDS \cite{arganids} uses Adversarially Regularised Graph Autoencoder (ARGA) \cite{pan2019adversariallyregularizedgraphautoencoder} GAE to learn node embeddings and then uses a downstream supervised Machine Learning classifier.

As mentioned in the Introduction, the main limitation of using supervised learning is the reliance on limited and possibly outdated datasets \cite{PINTO2025111177} due to the increased cost of labelling network data \cite{Braun_2024}. This cannot easily scale for a real-life implementation, leading to the need for unsupervised approaches.

When it comes to unsupervised approaches, there is much less related work. Anomal-E is a state of the art network anomaly detection algorithm that implements a GAE to learn meaningful embeddings of a graph \cite{Caville_2022}. Anomal-E implements E-GraphSAGE. A modified Graph Infomax version is then used to train the model encoder. The resulting embeddings are then fed to a downstream estimator used for anomaly detection. Although this approach is effective, the pipeline adds more complexity through additional components and tuning options. As such, we compare primarily unsupervised approaches, as we assume that there are no labels available during training.

\section{Methodology}
\label{sec:method}

In this section, we first describe the dataset construction and graph representation. We then focus on describing our methodology and how we have created, trained, and evaluated AutoGraphAD. Finally, we explain the anomaly scoring and evaluation.

\subsection{Dataset}

For this work, we used the UNSW-NB15 dataset \cite{7348942}. We chose UNSW-NB15 for the AutoGraphAD evaluation, as it is a widely used dataset with clearly labelled modern attacks. Thus, our evaluation is better suited for the current environment. The data set has been generated from the raw PCAP files using NFStream \cite{AOUINI2022108719}, which outputs flows with NetFlow V9 compatible features \cite{claise2004cisco}. To create the labels, we used the UNSW-NB15 generated labels \cite{7348942} and matched them to the generated flows.

The goal of this paper is to explore and showcase our model's ability to detect anomalies even when the training data are contaminated. To achieve this, we have created different training parts with different levels of contamination similar to Anomal-E \cite{Caville_2022}. The main difference between Anomal-E's approach and ours is that Anomal-E uses 10\% of the data set to create a single large graph for training, and the same is repeated for testing.

We use time windows to detect attacks and simulate real-life deployments. To achieve a fair distribution of attacks, we use stratified data split to see whether or not contaminated attacks are equally distributed. For 0\% contamination, we remove all the graphs with attacks, for contamination, we use a stratified split of the dataset, and for 5.7\% we downsample the benign graphs to  of their original number.

We allocate 70\% of the data for training, 20\% for evaluation, and 10\% for testing. For the time windows, we chose to create non-overlapping graphs that capture 180 seconds of traffic. For categorical features, which in our case is the IP version and protocols, we employ One-Hot encoding. In the case of the protocols, we use Rare Labelling from the library Feature Engine \cite{Galli2021} to reduce the dimensionality of the data. For continuous features, we use L2 row-wise normalisation.

The graph is made up of IP nodes that contain placeholder values used for message passing and connection nodes that contain flow features. The graph is undirected.

\begin{figure*}[!t]
\centering
\includegraphics[width=0.8\linewidth]{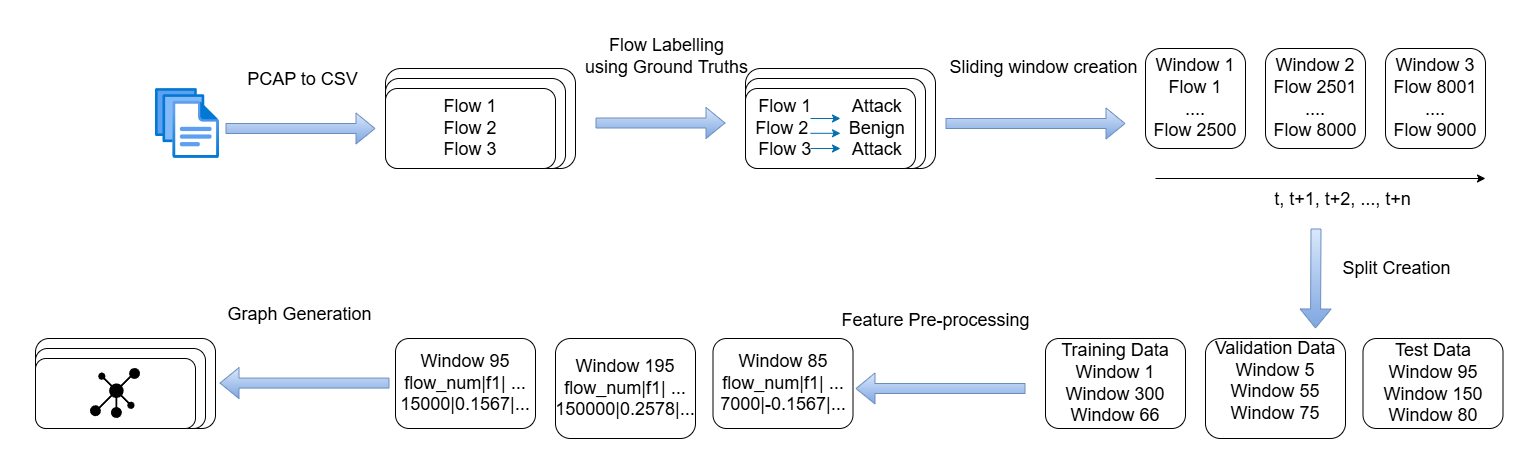}

\includegraphics[width=0.8\linewidth]{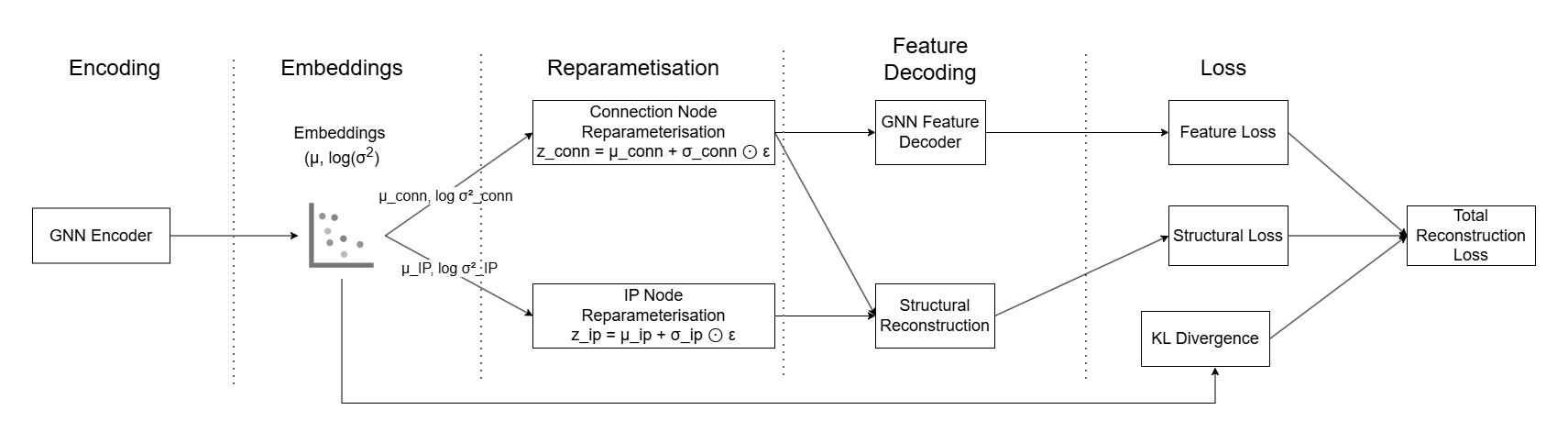}
\caption{The proposed Architecture of AutoGraphAD in a training setting. AutoGraphAD mainly focuses on the reconstruction and use of the connection nodes as the IP nodes have placeholder values. The pipeline starts with the encoder that generates the latent space embeddings that are then reparametised to create the embeddings that are then used for structure and feature reconstruction.
The GNN encoder and decoder as well as the losses can be switched out with every other GNN algorithm. In this paper, we are employing GraphSAGE as the encoder and decoder algorithm.}
    \label{fig:gave-architecture}
\end{figure*}

\subsection{Model Architecture}

Our proposed architecture builds on the original VGAE \cite{kipf2016variationalgraphautoencoders} and extends it by incorporating elements from GraphMAE \cite{GraphMAE} and HGMAE \cite{tian2023heterogeneousgraphmaskedautoencoders}. Our architecture is designed to reconstruct both the node features and the graph structure during training. The main upgrades introduced in our design are as follows.

We are using the same masking component as used in GraphMAE \cite{GraphMAE} and HGMAE \cite{tian2023heterogeneousgraphmaskedautoencoders}, but we only mask the original graph nodes. We made this decision because during the evaluation of AutoGraphAD there was a negligible improvement when using double masking.

We incorporated random edge drop during training to force the model to learn non-trivial relationships between nodes \cite{tian2023heterogeneousgraphmaskedautoencoders}. We have also incorporated negative edge sampling as a contrastive learning objective to help the autoencoder to better distinguish between real and fake edges.

For structural reconstruction, we use an enhanced Dot Product. The enhanced Dot Product uses learnable weights to better capture the structural dependencies of the graphs. This is because a heterogeneous graph can be split into smaller bipartite graphs \cite{Shi2022}. This allows us to assign a learnable weight for each of the edge types and better capture that individual relationship. For the structural reconstruction error, we use Binary Cross entropy loss as seen in the original VGAE paper \cite{kipf2016variationalgraphautoencoders}.

\begin{equation}
    A' = Z \cdot (W*Z^T) 
\end{equation}

Regarding feature reconstruction, a GNN model has been employed to decode only the connection node features. A GNN was chosen as it is the most capable of decoding the embeddings correctly. GNNs as a decoder are used in both HGMAE and GraphMAE \cite{tian2023heterogeneousgraphmaskedautoencoders, GraphMAE}, but in these approaches they have focused only on decoding masked embeddings. In this architecture, all of the connection nodes are decoded.

Regarding the feature reconstruction errors, we have created two main variants, one using the Cosine Embedding Error and one using the Mean Squared Error (MSE). The Cosine Embedding Error approach has been inspired by HGMAE \cite{tian2023heterogeneousgraphmaskedautoencoders} and GraphMAE \cite{GraphMAE}, while the MSE approach was chosen to help generate meaningful embeddings that are closest to the actual values. The KL divergence loss used in our model is the same as in the original VGAE paper \cite{kipf2016variationalgraphautoencoders}.

For the total reconstruction loss, we combine the feature reconstruction loss with the structural reconstruction loss and KL divergence. Each component of the total reconstruction loss, with the exception of the KL divergence, is weighted by a user-defined value. The model architecture can be found in Figure \ref{fig:gave-architecture} and the total loss is depicted in Equation \ref{ours-total-loss}.

\begin{equation}
    L_{Total} = \alpha*L_{Struct} + \beta*L_{Feat} + KL
    \label{ours-total-loss}
\end{equation}

The model was trained with 100 maximum allowable epochs. We used an early stopping mechanism during the training to prevent overfitting and expenditure of the computational resources. Furthermore, we saved the model's weights at the point that the model achieved its best performance.

\subsection{Anomaly Detection}

For anomaly detection, we calculate the anomaly scores of the connection nodes. The anomaly scores are based on the calculation of the reconstruction errors with the KL term. The idea is that the anomaly score will be lower for normal nodes and higher for anomalous nodes. We use a percentile-based threshold, where a node is classified as anomalous if its score exceeds the p-th percentile; $p$ is selected offline as a threshold hyperparameter. For our score, we use the following equation.

\begin{equation}
    Score = \alpha * L_{Feat} + \beta * L_{Struct} + \gamma * KL
\end{equation}

The equation works by using user-assigned weights to weigh the importance of each loss in the anomaly score. This allows us to use different weights for different settings and adapt to the changing nature of the network. We employ Robust Scaling \cite{https://doi.org/10.1002/widm.1236} of all reconstruction errors and KL Divergence that are used for anomaly scores to ensure that no value dominates the scale. Our entire approach runs in PyTorch and on the GPU.

\subsection{Evaluating Inference Performance}

To evaluate the performance of each method, we have chosen to measure the training and inference times. For timing, we used the Python built-in library, time.

We do not count the parts where data transfer takes place to make the comparisons more equal. Moreover, in the case of AutoGraphAD, we can time each individual pass by generating the Anomaly Score. However, we cannot time each pass in the threshold setting, threshold inference timing, and Anomal-E's downstream estimator training and inference timing. In that form, we get the average pass time by dividing the total by all of the passes. 

\subsection{Libraries Used}

For AutoGraphAD, we used PyTorch, PyTorch Geometric \cite{fey2019fastgraphrepresentationlearning} and DGL \cite{wang2020deepgraphlibrarygraphcentric} for the GNNs. For faster training, we employed PyTorch Lightning \cite{Falcon_PyTorch_Lightning_2019}. NumPy and PyOD \cite{zhao2019pyod} are used for anomaly detection. For data pre-processing, window creation, and running downstream tasks for Anomal-E, we used Scikit-learn \cite{scikit-learn} and Pandas.

\section{Experimental Results}

\subsection{Experimental setup}

To evaluate AutoGraphAD, we used UNSW-NB15 in both AutoGraphAD and Anomal-E. For Anomal-E, we used an implementation found in a GitHub repository\footnote{https://github.com/waimorris/Anomal-E}. We chose to only evaluate Anomal-E as it is a pure Anomaly Detection paper.

Overall, we trained the models in the different contamination levels followed by the anomaly detection estimators' grid search for Anomal-E and weight search for the Anomaly Score in the VGAE implementation. During the hyperparameter search, the model weights were frozen. Labels are used only for offline evaluation and score-weight selection for anomaly detection.

To run our experiments, we used a server with an Nvidia RTX 3090 with 24 GB of VRAM. The CPU of the server is an AMD Ryzen 3950X, 16-Core Processor. The server came with 64 GB of RAM and its operating system was Ubuntu 22.04.4 LTS.

\subsection{VGAE Variants}

We exhaustively tested multiple different VGAE variants. Our base architecture is node masking and edge dropping to force the variational autoencoders to learn non-trivial node relationships. The variants that we evaluated are the following: $(1)$ Base architecture using MSE for node feature reconstruction, $(2)$ Base architecture using Cosine Embedding Error for node feature reconstruction, $(3)$ Regularised Base Architecture, $(4)$ Split Base Architecture and $(5)$ Multiple Sampling Base Architecture.

We evaluated each version using 1 and 2 layer encoders and decoders, as well as using KL annealing. The implementation of all our variants are available in the accompanying code repository.

\begin{figure}[ht!]
    \centering
    \includegraphics[width=0.9\linewidth]{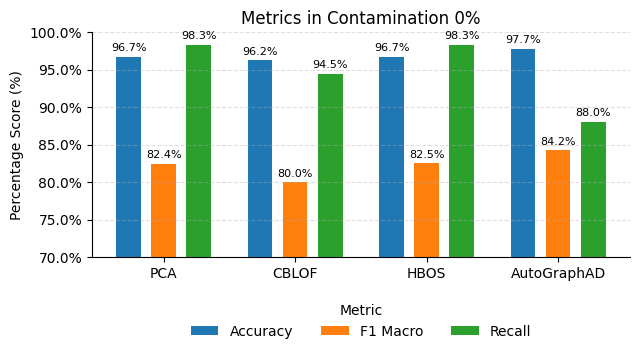}
    \caption{Performance metrics in 0\% training dataset.}
    \label{fig:0cont-metric}
\end{figure}

\subsection{Training Hyperparameters}
 
For training the Anomal-E dataset we used the hyperparameters present in the GitHub's repository. These were verified to be the same hyperparameters as those mentioned in the original paper \cite{Caville_2022}.

For AutoGraphAD, we have run multiple versions of our architecture. We have run different versions with different upgrades to evaluate which yields the best results. We relied on Edge Index accuracy to select the best version of our model.

\subsection{Anomaly Detection Hyperparameters}

Anomal-E requires a downstream detector to characterise attacks as anomalies or not. In the case of Anomal-E we chose to use the same algorithms as they used in their original paper \cite{Caville_2022}. For the Hyperparameter search, we used a more extensive search space, the grid can be found in our repository\footnote{\url{https://github.com/georgeani/AutoGraphAD/}}. Our Anomal-E implementation, including all variants evaluated, is available in our code repository. We used a Brute Force Hyperparameter search where we used the dataset labels to tune the parameters. We chose the results that yielded the best balance between F1-Score, Precision, and Recall.

\begin{figure}[ht!]
    \centering
    \includegraphics[width=0.9\linewidth]{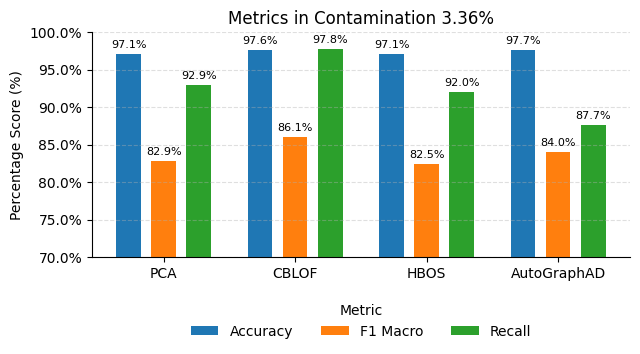}
    \caption{Performance metrics at 3.36\% contamination in the training dataset.}
    \label{fig:35cont-metrics}
\end{figure}

\begin{table*}[!t]
  \caption{Results of Anomal-E and AutoGraphAD}
  \label{tab:threecols}
  \centering
  \footnotesize
  \setlength{\tabcolsep}{8pt}
  \renewcommand{\arraystretch}{1.15}

  For AutoGraphAD, we have tried different settings to find the combination that yields the most reliable anomaly scores. A detailed summary of the different parameters that we have evaluated can be found in our repository.

  \begin{tabular*}{\textwidth}{@{\extracolsep{\fill}} l ccc ccc ccc}
    \toprule
      & \multicolumn{3}{c}{0\% Contamination} & \multicolumn{3}{c}{3.36\% Contamination} & \multicolumn{3}{c}{5.76\% Contamination} \\
    \cmidrule(lr){2-4}\cmidrule(lr){5-7}\cmidrule(lr){8-10}
      Metrics & Accuracy & F1 & Recall & Accuracy & F1 & Recall & Accuracy & F1 & Recall \\
    \midrule
      PCA (Anomal-E)
      & 96.65\% & 82.39\% & 98.27\%
      & 97.3\% & 82,56\% & 95.81\%
      & 97.13\% & 84.24\% & 92.95\% \\
      
      CBLOF (Anomal-E)
      & 96.21\% & 79.96\% & 94.46\%
      & 95.11\% & 77.51\% & 97.47\%
      & 97.63\% & 86.05\% & 97.76\% \\
      
      HBOS (Anomal-E)
      & 96.68\% & 82.5\% & 98.28\%
      & 97.37\% & 84.73\% & 96.61\%
      & 97.1\% & 82.48\% & 91.97\% \\
      
      \textbf{Anomaly Score (AutoGraphAD)}
      & 97.69\% & 84.23\% & 87.98\% 
      & 97.67\% & 84.03\% & 87.67\%
      & 97.6\% & 83.36\% & 86.42\% \\
    \bottomrule
  \end{tabular*}
\end{table*}

\subsection{Results and interpretation}

For the results, we have chosen the best of our variants for each level of contamination and different weights for the anomaly score. 
The metrics used are Accuracy, F1 Macro and Recall Macro. These metrics were chosen as an accurate descriptor of the model's performance and because Macro treats each class as equal \cite{du2023skewsensitiveevaluationframeworkimbalanced}, making them ideal for evaluating the model in a highly unbalanced dataset like UNSW-NB15 \cite{7348942}. At 0\% and 3.36\% contamination, we used our regularised variant, and at 5.76\% we used the multiple sampling variant.

\begin{figure}[ht!]
    \centering
    \includegraphics[width=0.9\linewidth]{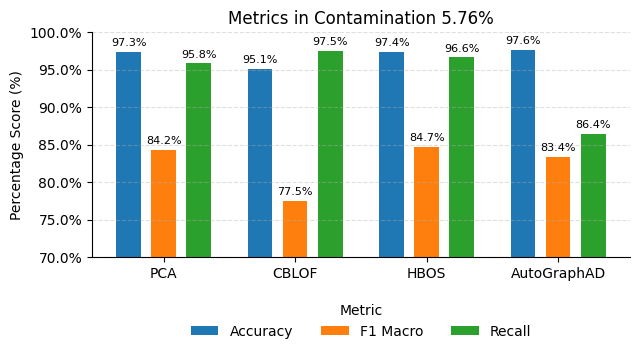}
    \caption{Performance metrics of all the approaches at 5.76\% contamination.}
    \label{fig:57cont-metrics}
\end{figure}

Figures \ref{fig:0cont-metric}, \ref{fig:35cont-metrics}, and \ref{fig:57cont-metrics} show the results of our experimental evaluation for the different levels of contamination.
AutoGraphAD yields very good results, comparable to those of Anomal-E. It should be noted that AutoGraphAD does not rely on any downstream anomaly detection algorithms such as Anomal-E, making it ideal for real-life deployment, as it can be tuned and adjusted more easily rather than requiring training on the new dataset.

Our results at 0\% contamination exceed Anomal-E's, especially when it comes to accuracy and F1 results. Anomal-E seems to be performing better in Recall, but looking at the F1 score its Precision seems to be lower than AutoGraphAD, indicating a higher false positive count than our own approach. This continues at the 3.36\% contamination level, where only CBLOF seems to perform better than AutoGraphAD. In the case of 5.76\% contamination levels, again we achieve higher results with the exception of Recall against most of Anomal-E's estimators.

AutoGraphAD yields substantially better inference time (0.009892 s) compared to PCA (0.0286 s), CBLOF (0.0323 s) and HBOS(0.0869 s). This can also be observed in training: AutoGraphAD (0.011305 s), PCA (0.0885 s), CBLOF(0.0454 s), HBOS (0.1359 s). This can mainly be attributed to the fact that AutoGraphAD does not require a downstream detector as Anomal-E.

AutoGraphAD is capable of being more easily tuned by simply adjusting the weight parameters of the anomaly score, making alterations to the anomaly score generation on the fly. Unfortunately, this cannot truly be done in the case of the estimators as their characteristics are set during initialisation. Some of the estimator's thresholds may be manually adjusted \cite{zhao2019pyod} \cite{scikit-learn}, but this is also the case in AutoGraphAD. Finally, in the case of model retuning or retraining, Anomal-E's estimators need to be re-trained from the start as the PyOD's architecture does not support retuning \cite{zhao2019pyod}. Instead, AutoGraphAD is very easy to retune both the model and the anomaly score due to its design.

\section{Conclusion}

In conclusion, AutoGraphAD leverages recent advancements in the space of GVAEs and GAEs \cite{GraphMAE, tian2023heterogeneousgraphmaskedautoencoders, kipf2016variationalgraphautoencoders, pan2019adversariallyregularizedgraphautoencoder} to build a more practical approach for network anomaly detection. Our design is relatively simple and yields results comparable to those of the Anomal-E. The main advantage of AutoGraphAD is faster inference and detection times, a lower number of false positives, and easier and faster pipeline retuning than Anomal-E's approach.

Our future work will focus on evaluating the generalisation capabilities of our model across unseen network scenarios, including a broader range of datasets. Assessing the capacity of a model to operate in a different environment than those seen in training is also essential to ensure practical deployment in operational networks. 

\section*{Acknowledgements}

This work was supported by Grant PCI2023-145974-2 funded by MICIU/AEI/10.13039/501100011033 and cofunded by the European Union (GRAPHS4SEC project). This work is also supported by the Catalan Institution for Research and Advanced Studies (ICREA Academia). GitHub Copilot was used to help debug and develop the code. It served mainly as a search engine for what was causing the bugs and as a recommendation engine to check which libraries and their methods are the most suitable and optimal during development.

\balance
\bibliographystyle{IEEEtran}
\bibliography{bibliography}

\end{document}